\documentclass[preprint,aps,prb,showpacs,amsmath]{revtex4}
\usepackage{latexsym}
\usepackage{amssymb}
\usepackage{graphicx}
\usepackage{bm}% bold math
\usepackage[english]{babel}
\usepackage[latin1]{inputenc}
\usepackage{amsmath}
\usepackage{amsfonts}

\newcommand{\bq}{\begin{equation}}
\newcommand{\eq}{\end{equation}}
\newcommand{\bn}{\begin{eqnarray}}
\newcommand{\en}{\end{eqnarray}}

\begin{document}

\title{Magnetic-flux-controlled giant Fano factor for the coherent tunneling through a %%@
parallel double-quantum-dot}

\author{Bing Dong}
\affiliation{Department of Physics, Shanghai Jiaotong University, 1954 Huashan
Road, Shanghai 200030, China}
\author{X. L. Lei}
\affiliation{Department of Physics, Shanghai Jiaotong University, 1954 Huashan
Road, Shanghai 200030, China}

\begin{abstract}

We report our studies of zero-frequency shot noise in tunneling through a parallel-coupled %%@
quantum dot interferometer by employing number-resolved quantum rate equations. We show that %%@
the combination of quantum interference effect between two pathways and strong Coulomb %%@
repulsion could result in a giant Fano factor, which is controllable by tuning the enclosed %%@
magnetic flux.

\end{abstract}


\maketitle

Quantum shot noise in nanosystem far from equilibrium has become currently active issue since %%@
it characterizes the degree of correlation between
charge transport events, which can not be obtained by measuring mean current %%@
alone.\cite{Blanter} In particular, it is specially interesting to search for mesoscopic %%@
devices having super-Poissonian noise %%@
feature.\cite{Jong,Sukhorukov,Cottet,Wacker,Nauen,Safonov,Djuric} For instance, a single %%@
quantum dot (QD) with multilevel connected to two leads was recently reported to show %%@
enhanced noise by analyzing its full counting statistic, in which quantum interference %%@
between multi-states was not taken into account.\cite{Belzig}   

In this letter, we study the shot noise of first-order resonant tunneling through a two-level %%@
system, a two parallel-coupled QDs, under an extremely large bias-voltage with emphasis on %%@
the quantum interference effect. The theoretical model is depicted in Fig.~1, in which both %%@
of two QDs 1 and 2 are connected to the left and right leads via the tunneling matrix element %%@
$V_{\eta j}$ ($\eta=\{L,R\}$ and $j=\{1,2\}$) describing the coupling between QD $j$ and lead %%@
$\eta$. For simplicity, we assume $V_{L1}=V_{R2}$ and $V_{L2}=V_{R1}$, and the strength of %%@
tunneling $\Gamma=2\pi \varrho_0 |V_{L1(R2)}|^2$ ($\varrho_0$ is the flat density of states %%@
of the leads) and $\Gamma'=2\pi \varrho_0 |V_{L2(R1)}|^2$ being constant in the wide band %%@
limit. 
Here we also assume that only one spinless energy level $\varepsilon_j$ in each dot is %%@
involved in transport. $\Omega$ denotes dot-dot hopping and a magnetic flux $\varphi\equiv %%@
2\pi \Phi/\Phi_0$ ($\Phi_0\equiv hc/e$ is the magnetic flux quantum) is applied to penetrate %%@
the area enclosed by the two tunneling pathways. Note that this QD Aharonov-Bohm %%@
interferometer has been realized in recent experiments.\cite{Holleitner,ChenJ}     

%\begin{figure}[htb]
%\includegraphics[height=4cm,width=8cm]{fig1}
%\caption{Schematic diagram for the coherent resonant tunneling through a parallel-coupled %%@
%double quantum dot Aharonov-Bohm interferometer.}
%\label{fig1}
%\end{figure}

Under the limit of sufficiently large bias-voltage ($V\gg \Gamma,\Gamma',\Omega$), electronic %%@
tunneling through this system in first-order picture can
be described by the quantum
rate equations (QREs) for the dynamic evolution of the reduced density matrix
elements of the coupled QDs, $\rho_{ab}(t)$ ($a,b=\{0, 1, 2, d\}$).\cite{Gurvitz,Dong1,Dong} 
The diagonal elements of the reduced density matrix, $\rho_{aa}$, give the occupation %%@
probabilities of the
states of the dots, namely: $\rho_{00}$($\rho_{dd}$) is the probability of
finding both dots unoccupied(occupied), $\rho_{11}$ and $\rho_{22}$ are the probabilities of %%@
finding dot 1 and dot 2 occupied, respectively; while the off-diagonal element %%@
$\rho_{12}=\rho_{21}^{\ast}$ describe the coherent
superposition state between two QDs. For the purpose of evaluating the noise spectrum, we %%@
introduce the number-resolved density matrices $\rho_{ab}^{(n)}(t)$, meaning that the system %%@
is in the electronic state $|a\rangle$ ($a=b$) or in quantum superposition state ($a\neq b$) %%@
at time $t$ and meanwhile total $n$ electrons are counted to transfer into the right lead by %%@
time $t$.\cite{Gurvitz} Obviously, $\rho_{ab}(t)=\sum_{n} \rho_{ab}^{(n)}(t)$ and the %%@
resulting
number-resolved QREs at zero temperature and large bias-voltage in the case of infinite %%@
inter-dot Coulomb interaction are:
\begin{subequations}
\label{rateq1} 
\bn
\dot{\rho}_{00}^{(n)}&=& \Gamma' \rho_{11}^{(n-1)} + \Gamma \rho_{22}^{(n-1)} - (\Gamma + %%@
\Gamma') \rho_{00}^{(n)} \cr
&& + \kappa\sqrt{\Gamma\Gamma'} (e^{-i\varphi/2} \rho_{12}^{(n-1)} + e^{i\varphi/2} %%@
\rho_{21}^{(n-1)}), \label{rc01} \\
\dot{\rho}_{11}^{(n)} &=& \Gamma \rho_{00}^{(n)} - \Gamma' \rho_{11}^{(n)} +i \Omega ( %%@
\rho_{12}^{(n)}-\rho_{21}^{(n)}) \cr
&& -\frac{\kappa}{2} \sqrt{\Gamma\Gamma'} (e^{-i\varphi/2} \rho_{12}^{(n)} + e^{i\varphi/2} %%@
\rho_{21}^{(n)}), \label{rc11} \\
\dot{\rho}_{22}^{(n)} &=& \Gamma' \rho_{00}^{(n)} - \Gamma \rho_{22}^{(n)} -i\Omega %%@
(\rho_{12}^{(n)}-\rho_{21}^{(n)}) \cr
&& -\frac{\kappa}{2} \sqrt{\Gamma\Gamma'} (e^{-i\varphi/2} \rho_{12}^{(n)} + e^{i\varphi/2} %%@
\rho_{21}^{(n)}), \label{rc21} \\
\dot{\rho}_{12}^{(n)} &=& i\Omega (\rho_{11}^{(n)}-\rho_{22}^{(n)}) - \frac{1}{2} (\Gamma + %%@
\Gamma') \rho_{12}^{(n)} \cr
&& \hspace{-1cm} + \sqrt{\Gamma\Gamma'} e^{-i\varphi/2} \rho_{00}^{(n)} - \frac{\kappa}{2} %%@
\sqrt{\Gamma\Gamma'} e^{i\varphi/2} (\rho_{11}^{(n)} + \rho_{22}^{(n)}),\label{rc31} 
\en
\end{subequations}
together with the normalization relation
$\rho_{00}+\rho_{11}+\rho_{22}=1$. The adjoint equation of Eq.~(\ref{rc31}) gives
the equation of motion for $\rho_{21}^{(n)}$. The parameter $\kappa\sqrt{\Gamma\Gamma'}$ %%@
describes the interference in tunneling events through different pathways, in which $\kappa$ %%@
is artificially introduced to describe decoherence [$\kappa=0(1)$ means full noninterference %%@
(interference) between two pathways].\cite{Shahbazyan,Gurvitz1} From these number-resolved %%@
QREs, we can readily deduce the usual QREs for the reduced density matrix elements, ${\bm %%@
\rho}(t)=(\rho_{00},\rho_{11},\rho_{22},\rho_{12},\rho_{21})^{T}$, as: $\dot {\bm %%@
\rho}(t)={\cal M}{\bm \rho}(t)$, where ${\cal M}$ can be easily read from %%@
Eqs.~(\ref{rc01})-(\ref{rc31}). 

The current flowing through the system can be evaluated by the time change rate of electron %%@
number in the right lead (we use $\hbar=e=1$)
\bq
I = \dot N_{R}(t) =\frac{d}{dt} \sum_{n} nP_n(t){\Big |}_{t\rightarrow\infty}, \label{I}
\eq
where $P_n(t) = \rho_{00}^{(n)}(t)+ \rho_{11}^{(n)}(t)+ \rho_{22}^{(n)}(t)$ 
is the total probability of transferring $n$ electrons into the right lead by time $t$. %%@
According to the MacDonald's formula for shot noise,\cite{MacDonald} the zero-frequency shot %%@
noise with respect to the right lead is defined by $P_n(t)$ as well:\cite{Chen,Elattari}
\bq
S(0)=\frac{d}{dt} \left [ \sum_{n} n^2 P_n(t) - (t I)^2 \right ] {\Big %%@
|}_{t\rightarrow\infty}.
\eq 
With the help of Eqs.~(\ref{rateq1}), the current and shot noise can be written as
\bn
I &=& \left [ \Gamma' \rho_{11} + \Gamma \rho_{22} + \kappa\sqrt{\Gamma\Gamma'} \right. \cr
&& \left. \times (e^{-i\varphi/2} \rho_{12} + e^{i\varphi/2} \rho_{21}) \right ]{\big %%@
|}_{t\rightarrow\infty}, \label{current} \\
S(0) &=& \left \{ \Gamma' (2G_{11} + \rho_{11}) + \Gamma (2G_{22} + \rho_{22}) + %%@
\kappa\sqrt{\Gamma\Gamma'} \right. \cr
&& \hspace{-1.5cm} \left. \times\left [ e^{-i\varphi/2} (2G_{12}+ \rho_{12}) + e^{i\varphi/2} %%@
(2G_{21}+ \rho_{21})\right ]\right \}{\big |}_{t\rightarrow\infty}, \label{noise}
\en 
with the generating function $G_{ab}(t)$ defined as
\bq
G_{ab}(t)=\sum_{n} n \rho_{ab}^{(n)}(t).
\eq
Employing Eqs.~(\ref{rateq1}), the equations of motion for ${\bm %%@
G}(t)=(G_{00},G_{11},G_{22},G_{12},G_{21})^{T}$ are explicitly obtained in matrix form: %%@
$\dot{\bm G}(t)={\cal M}{\bm G}(t) + {\cal G}{\bm \rho}(t)$ with ${\cal G}_{12}=\Gamma'$, %%@
${\cal G}_{13}=\Gamma$, ${\cal G}_{14(5)}=\kappa \sqrt{\Gamma\Gamma'} e^{\mp i\varphi/2}$, %%@
and all other elements being zero.
Applying Laplace transform to these equations yields
\bq
{\bm G}(s) = (s {\bm I}-{\cal M})^{-1} {\cal G} {\bm \rho}(s),
\eq
where ${\bm \rho}(s)$ is readily obtained by performing Laplace transform on its equations of %%@
motion with the initial condition ${\bm \rho}(0)={\bm \rho}_{st}$ (${\bm \rho}_{st}$ denotes %%@
the stationary solution of the QREs). Due to inherent long-time stability of the physics %%@
system under investigation, all real parts of nonzero poles of ${\bm \rho}(s)$ and ${\bm %%@
G}(s)$ are negative definite. Consequently, the large-$t$ behavior of the auxiliary functions %%@
is entirely determined by the divergent terms of the partial fraction expansions of ${\bm %%@
\rho}(s)$ and ${\bm G}(s)$ at $s\rightarrow 0$. 

Finally, we arrive at analytical expressions for the current from Eq.~(\ref{current}) %%@
($x=\Omega/\Gamma$ and $\gamma=\Gamma'/\Gamma$):
\bq
I_1=\frac{4x^2 (\gamma+1)} {(\gamma+1)^2+12 x^2} \Gamma,
\eq
for the fully interferential case ($\kappa=1$) and
\bq
I_0=\frac{(\gamma+4x^2) (\gamma+1)} {(\gamma+1)^2 + 12x^2} \Gamma,
\eq
for the fully noninterferential case ($\kappa=0$). It is found that the current is %%@
independent of magnetic flux due to $\rho_{00}$ being constant function of magnetic flux in %%@
the strong Coulomb blockade limit. However, the derivation is difficult for the %%@
zero-frequency shot noise Eq.~(\ref{noise}). We obtain analytical expressions only for %%@
several magnetic fluxes: if $\varphi=0$ and $2\pi$, we have    
\bn
S_1(0) &=& 4x^2 \Gamma (\gamma+1)[(80\gamma^2+352\gamma+80)x^4 \cr
&& +(-8\gamma^4+160\gamma^3+336 \gamma^2+160\gamma-8)x^2 \cr 
&& +\gamma^6+10\gamma^5+31\gamma^4+44\gamma^3+31\gamma^2+10\gamma+1] \cr
&& \times (\gamma-1)^{-2} [(\gamma+1)^2+12 x^2]^{-3}, \label{sn0}
\en
at $\kappa=1$;
while if $\varphi=\pm \pi$ and $\kappa=1$, we obtain
\bn
S_1(0) &=& 4x^2 \Gamma (\gamma+1) [80 x^4-(8\gamma^2 +16\gamma +8)x^2+\gamma^4 \cr
&& +4\gamma^3+6\gamma^2+4\gamma+1] [(\gamma+1)^2+12 x^2]^{-3}. \label{sn1}
\en
Let's discuss two special situations. When $\Gamma'=0$ ($\gamma=0$), the system reduces to %%@
series-coupled QDs, and correspondingly, the current and shot noise become 
\bq
I = \frac{4x^2 }{1+12 x^2} \Gamma, \,\,\, S(0) = \frac{4x^2 (80 x^4 - 8x^2 +1)}{(1+12 x^2)^3} %%@
\Gamma,
\eq
irrespective of $\kappa$, which are identical to the previous results.\cite{Elattari}  
For a completely symmetric interferometer $\Gamma'=\Gamma$ ($\gamma=1$), the current is
\bq
I= \left \{
\begin{array}{ccl}
\frac{2}{3} \Gamma, & \, & \text{if $\varphi=0$ and $2\pi$,} \\
\frac{8x^2-2\kappa^2+2}{3+12x^2+2\kappa-\kappa^2} \Gamma, & \, & \text{if $\varphi=\pm \pi$}.
\end{array}
\right.
\eq
While the zero-frequency shot noise becomes
\bq
S_1(0)= \left \{ 
\begin{array}{ccl}
\frac{1}{3} \Gamma, & \, & \text{if $\varphi=0$ and $2\pi$,} \\ 
\frac{2 x^2(5 x^4-2 x^2+1)}{(1+3 x^2)^3} \Gamma, & \, & \text{if $\varphi=\pm \pi$,} 
\end{array}
\right. \label{inf}
\eq
in the case of full coherence.
By contrast, for the full noninterference case, we have a constant shot noise:
\bq
S_0(0)= \frac{10}{27}\Gamma.
\eq

%\begin{figure}[htb]
%\includegraphics[height=8cm,width=8cm]{snop-c}
%\caption{(a) Current $I$, zero-frequency shot noise $S(0)$ (with unit $\Gamma$) and (b) Fano %%@
%factor $F=S(0)/I$ vs $\varphi$ for various values of $\Gamma'/\Gamma$ with %%@
%$\Omega/\Gamma=1/2$ and $\kappa=1$. Inset: $F$ vs $\kappa$.}
%\end{figure}
  
For other values of magnetic flux, we have to resort to numerical calculation for shot noise. %%@
In Fig.~2, we plot the calculated current $I$, zero-frequency shot noise $S(0)$, and Fano %%@
factor $F=S(0)/I$ with $x=1/2$ and $\kappa=1$ as functions of magnetic flux for different %%@
coupling $\gamma$ of the additional pathway in the case of infinite interdot Coulomb %%@
repulsion.
Our results explicitly show that (1) shot noise and Fano factor are periodic functions of %%@
magnetic flux with a period of $2\pi$; (2) shot noise is significantly enhanced due to the %%@
destructive quantum interference effect around $\varphi=0$ and $\pm 2n\pi$ ($n$ is an %%@
integer), leading to a giant Fano factor up to $10^2$ at $\gamma=0.8$ for full coherence %%@
$\kappa=1$, while the shot noise is always sub-Poissonian for the system without the %%@
additional tunneling path $\gamma=0$; (3) the instructive quantum interference at %%@
$\varphi=\pm n\pi$ ($n$ is an odd integer) suppresses the Fano factor lower than unit, %%@
exhibiting sub-Poissonian noise; (4) the inset clearly shows that the giant Fano factor %%@
crucially depends on the degree of interference $\kappa$. This magnetic-flux-dependent noise %%@
property was also reported in the cotunneling transport through a %%@
parallel-CQD.\cite{Sukhorukov}     

However, the shot noise is always sub-Poissonian in the case of no interdot Coulomb %%@
repulsion. In this case, the associated QREs become ($\kappa=1$)
\begin{subequations}
\label{rateq2} 
\bn
\dot{\rho}_{11}^{(n)} &=& \Gamma \rho_{00}^{(n)} - 2 \Gamma' \rho_{11}^{(n)} +\Gamma %%@
\rho_{dd}^{(n-1)} + i \Omega ( \rho_{12}^{(n)}-\rho_{21}^{(n)}) \cr
&& -\frac{1}{2} \sqrt{\Gamma\Gamma'} (e^{-i\varphi/2} - e^{i\varphi/2}) (\rho_{12}^{(n)}- %%@
\rho_{21}^{(n)}), \label{nrc11} \\
\dot{\rho}_{22}^{(n)} &=& \Gamma' \rho_{00}^{(n)} - 2\Gamma \rho_{22}^{(n)} + \Gamma' %%@
\rho_{dd}^{(n-1)} -i\Omega (\rho_{12}^{(n)}-\rho_{21}^{(n)}) \cr
&& -\frac{1}{2} \sqrt{\Gamma\Gamma'} (e^{-i\varphi/2} - e^{i\varphi/2}) (\rho_{12}^{(n)}- %%@
\rho_{21}^{(n)}), \label{nrc21} \\
\dot{\rho}_{dd}^{(n)} &=& \Gamma' \rho_{11}^{(n)} + \Gamma \rho_{22}^{(n)} - (\Gamma+ %%@
\Gamma') \rho_{dd}^{(n)} \cr
&& - \sqrt{\Gamma\Gamma'}e^{i\varphi /2} \rho_{12}^{(n)} - \sqrt{\Gamma\Gamma'}e^{-i\varphi %%@
/2} \rho_{21}^{(n)}, \\
\dot{\rho}_{12}^{(n)} &=& i\Omega (\rho_{11}^{(n)}-\rho_{22}^{(n)}) - (\Gamma + \Gamma') %%@
\rho_{12}^{(n)} \cr
&& + \sqrt{\Gamma\Gamma'} e^{-i\varphi/2} \rho_{00}^{(n)} - \sqrt{\Gamma\Gamma'} %%@
e^{i\varphi/2} \rho_{dd}^{(n-1)} \cr
&& - \frac{1}{2} \sqrt{\Gamma\Gamma'} (e^{i\varphi/2}- e^{-i\varphi/2} ) (\rho_{11}^{(n)} + %%@
\rho_{22}^{(n)}), \label{nrc31} 
\en
\end{subequations}
together with the normalization relation $\rho_{00}+\rho_{11}+\rho_{22} + \rho_{dd}=1$ %%@
[Equation for $\rho_{00}^{(n)}$ is the same as Eq.~(\ref{rc01})], which are quoted directly %%@
from our previous derivation using nonequilibrium Green's function.\cite{Dong} Along the same %%@
scheme as above indicated, we obtain the following results: (1) for $\Gamma'=0$ (the %%@
series-coupled QDs), 
\bq
I=\frac{2x^2}{1+4 x^2} \Gamma, \,\,\, S(0)=\frac{2x^2 (8 x^4 -2 x^2 +1)}{(1+ 4 x^2)^3} %%@
\Gamma,
\eq
which are also identical to the previous results;\cite{Elattari} and (2) for $\Gamma'=\Gamma$ %%@
(a completely symmetric interferometer),
\bq
\begin{array}{ccl}
I=\Gamma, & S(0)=\Gamma/2, \,\,\,\,\,& \text{if $\varphi=0$ and $2\pi$,} \\ 
I= \frac{x^2 }{1+x^2} \Gamma, & S(0)=\frac{x^2 (x^4 - x^2+ 2)}{2(1+x^2)^3} \Gamma, %%@
\,\,\,\,\,& \text{if $\varphi=\pm \pi$.} 
\end{array}
\eq

In conclusion, we have analyzed the shot noise properties of
resonant tunneling through parallel-coupled QD interferometer at extremely large %%@
bias-voltage. Our analytic and numerical results predict that a giant Fano factor can be %%@
found in this system due to the combination of quantum interference effect between two %%@
tunneling paths and strong Coulomb blockade, and a super-Poissonian--sub-Poissonian %%@
transition of the shot noise occurs by tuning the enclosed magnetic flux to change quantum %%@
interference pattern. 

This work was supported by Projects of the National Science Foundation of China, the Shanghai %%@
Municipal Commission of Science and Technology, the Shanghai Pujiang Program, and NCET.

\newpage

\centerline{\Large Figure Caption}

\vspace{2cm}

\noindent FIG.1: Schematic diagram for the coherent resonant tunneling through a %%@
parallel-coupled double quantum dot Aharonov-Bohm interferometer.

\vspace{1cm} 
\noindent FIG.2: (a) Current $I$, zero-frequency shot noise $S(0)$ (with unit $\Gamma$) and %%@
(b) Fano factor $F=S(0)/I$ vs $\varphi$ for various values of $\Gamma'/\Gamma$ with %%@
$\Omega/\Gamma=1/2$ and $\kappa=1$. Inset: $F$ vs $\kappa$.

\newpage

\begin{figure}[htb]
\includegraphics[height=4cm,width=8cm]{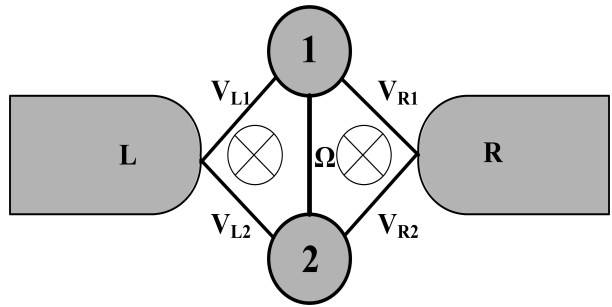}
\caption{Schematic diagram for the coherent resonant tunneling through a parallel-coupled %%@
double quantum dot Aharonov-Bohm interferometer.}
\label{fig1}
\end{figure}

\newpage

\begin{figure}[htb]
\includegraphics[height=8cm,width=8cm]{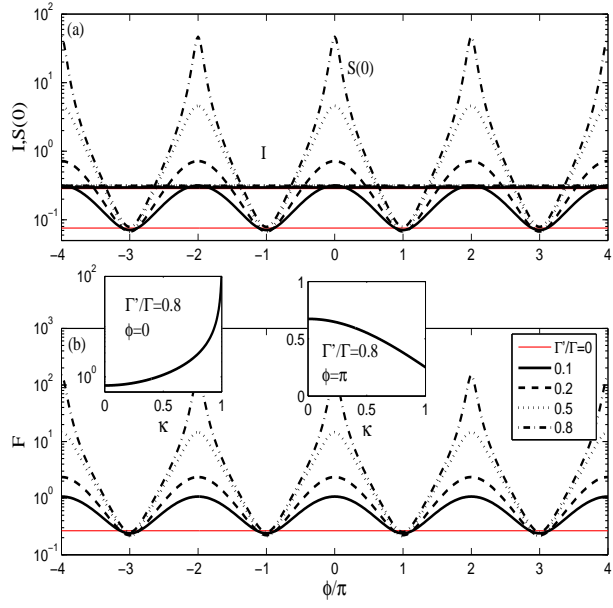}
\caption{(a) Current $I$, zero-frequency shot noise $S(0)$ (with unit $\Gamma$) and (b) Fano %%@
factor $F=S(0)/I$ vs $\varphi$ for various values of $\Gamma'/\Gamma$ with %%@
$\Omega/\Gamma=1/2$ and $\kappa=1$. Inset: $F$ vs $\kappa$.}
\end{figure}

\newpage

\end{document}